\documentclass[11pt]{article}
\usepackage{bm}
\usepackage{graphicx}
\begin{document}

\title{Neutrino Mass Sum-rule and Neutrinoless Double Beta Decay} 
\author{Asan Damanik\\ {\it Department of Physics Education, Sanata Dharma University}\\ {\it Kampus III USD Paingan Maguwoharjo Sleman Yogyakarta, Indonesia}\\{E-mail: asandamanik11@gmail.com}}
\date{}

\maketitle

\abstract{Neutrino mass sum-rule is a very important research subject from theoretical side because neutrino oscillation experiment only gave us two squared-mass differences and three mixing angles.  We review neutrino mass sum-rule in literature that have been reported by many authors and discuss its phenomenological implications especially on neutrino mass and neutrinoless double beta decay by plotting effective Majorana mass $\left<m_{ee}\right>$ as function of the lightest neutrino mass both for normal and inverted hierarchy by using the central values of reported mixing angles and reported squared-mass differences as input.}

\section{Introduction}
As we have already knew from the Standard Model of Particle Physics especially electroweak interaction model based on $SU(2)_{L}\times U(1)_{Y}$ gauge group, it is not possibble to obtain a neutrino mass term in the Lagrangian of electroweak interaction when neurino to be put as a Dirac particle.  But, if neutrino is a Majorana particle, then we can have a mass term in the Lagrangian which is given by
\begin{eqnarray}
L=\frac{1}{2}\nu_{L}C^{-1}M\nu_{L}+h,c.
\end{eqnarray}
where $\nu_{L}$ contains a three left-handed neutrino fileds, $C$ is the charge conjugation matrix, and $M$ is the Majorana mass matrix.

It was a very long time, before the neutrino oscillation phenomena was reported by the Superkamiokande collaboration in 1998 \cite{SuperK}, neutrino mass is assummed to be zero or approximately zero.  Unfortunately, the neutrino oscillations experiments only gave us the squared-mass difference between two neutrino flavors that undergo oscillations during its propagation in vacuum (two sqared-mass differences), and three mixing angles that cannot be used to determined the absolute value of neutrino mass and its hierarchy.  Another type of experiment that can be used to detect and determine the neutrino mass is  neutrinoless double beta decay experiment.  But, neutrinoless double beta decay experiment only give us upper bound of Majorana neutrino mass which is known as effective Majorana mass $\left<m_{ee}\right>$.  Thus, in order to determine neurino masses by using the experimental data as an input, we should seek another way or relation as an additional parameter that can be used to determine neutrino masses.  One of the relation that can be used to help us in determining the absolute value of neutrino mass is the relation that link all three neutrino masses which is know as {\it neutrino mass sum-rule}.  Neutrino mass sum-rule can also be interpreted geometrically as a triangle in complex plane, giving its area as a measure of CP violation \cite{Dorame}

In this paper, we review neutrino mass sum-rule that have already reported by many authors and discuss its phenomenological implications on neutrino masses, mass hierarchy, and effective Majorana mass.  The paper is organized as follow: in section 2 we review neutrino mass sum-rule that have already reported by many authors; in section 3 we discuss the penomenological implications of the neutrino mass sum rule on neutrino mass and effective Majorana mass especially for normal hierarchy.  Finally, the section 4 is devoted to conclusions.

\section{Brief review of neutrino mass sum-rule and neutrinoless double beta decay}
Neutrino mass sum rule is a relation among the neutrino masses $m_{1}, m_{2}, m_{3}$ which are known to be very small and it very useful for determining i. e. the hierarchy of neutrino mass whether it normal or inverted hierarchy, the absolute values of neutrino masses, and the effective neutrino mass $\left|m_{ee}\right|$  as measured in neutrinoless double beta decay when we use the experimental data of neutrino oscillation as input.  The importance of neutrino mass sum rule relation has already been stressed as well in e. g. Refs \cite {Chen,Barry,Spinrath}.  The neutrino mass sum rule, which can be obtained from several flavor models based on non-Abelian discrete symmetries, can be classified into four neutrino mass sum rules as one can reads in Ref. \cite{Dorame}
\begin{eqnarray}
\chi m_{2}+\xi m_{3}=m_{1},\label{A}\\
\frac{\chi}{m_{2}}+\frac{\xi}{m_{3}}=\frac{1}{m_{1}},\label{B}\\
\chi\sqrt{m_{2}}+\xi\sqrt{m_{3}}=\sqrt{m_{1}},\label{C}\\
\frac{\chi}{\sqrt{m_{2}}}+\frac{\xi}{\sqrt{m_{3}}}=\frac{1}{\sqrt{m_{1}}},\label{D}
\end{eqnarray}
where $\chi$ and $\xi$ are model dependent complex constants.  A sample of various neutrino mass sum rule and the groups generating them are summarized in \cite{King} and a summary table of the present neutrino mass sum rule in literatures can be found in \cite{Spinrath, Gehrlein}. As pointed out in \cite{Rojas} that the fisrt three mass sum rule, a classifcation of all models predicting tribimaximal (TBM) mixing which generates mass relations similar to the first three sum rule, but the last case is a completely new case.

By referring to Ref. \cite{Spinrath, Gehrlein}, in literature, we have known that there are twelve neutrino mass sum-rule according to the general mass sum-rule that can be parameterized as follow
\begin{eqnarray}
s(m_{1},m_{2},m_{3},c_{1},c_{2},\phi_{1},\phi_{2},d,\Delta_{\chi13},\Delta_{\chi 23})\equiv ~~~~~~~~~~~~~~~~~~~~~~~~~~~\nonumber\\c_{1}\left(m_{1}e^{-i\phi_{1}}\right)^{d}e^{i\Delta_{\chi 13}}+c_{2}\left(m_{2}e^{-i\phi_{2}}\right)^{d}e^{i\Delta_{\chi 23}}+m_{3}^{d}=0,\label{s}
\end{eqnarray}
where $\phi_{1}$ and $\phi_{2}$ are Majorana phases, and the quantities $c_{1},c_{2},d,\Delta_{\chi 13}$, and $\Delta_{\chi 23}$ are the parameters that characterize the sum-rule.

If we put $\phi_{1}=\phi_{2}=0$ and $\Delta_{\chi 13}=\Delta_{\chi 23}=0$, then Eq. (\ref{s}) reads
\begin{eqnarray}
c_{1}\left(m_{1}\right)^{d}+c_{2}\left(m_{2}\right)^{d}+m_{3}^{d}=0.\label{t}
\end{eqnarray}
It is apparent from Eq. (\ref{t}) the neutrino mass sum-rule of Eq. (\ref{A}) is easily obtained when we put $d=1$, $c_{1}=-\chi$, and $c_{2}=-\xi$.  To obtain Eq. (\ref{B}) from Eq. (\ref{t}) we should put  $d=-1$, $c_{1}=-\chi$, and $c_{2}=-\xi$, and Eq. (\ref{C}) is reproduced when we put $d=\frac{1}{2}$, $c_{1}=-\chi$, and $c_{2}=-\xi$.  Finally, Eq. (\ref{D}) will be obtained when we put  $d=-\frac{1}{2}$, $c_{1}=-\chi$, and $c_{2}=-\xi$ into Eq. (\ref{t}).  It is an important task to explain why there are four possible values of parameter $d$ and to decide what are the feasible value of $d$ which is in agreement with the experimental data.    
Another question is why there are twelve values for $\chi$ and $\xi$ that make possible twelve type of neutrino mass sum-rule as one can find in literature (see Table 1).  In this paper, we do not explain or answer the above questions, but we only evaluate and discuss the phenomenological implications of four types of neutrino mass sum-rule as shown in Eqs. (\ref{A})-(\ref{D}).

According to the experimental result of neutrino oscillation that the experimen only measure the squared mass difference, not absolute value of neutrino masses i. e. $\Delta m_{21}^{2}>0$ and $\Delta m_{31}^{2}>0$ or $\Delta m_{31}^{2}<0$, then we can have two possible hierarchies of neutrino mass i. e. normal hierarchy (NH) when $\Delta m_{21}^{2}>0$ and $\Delta m_{31}^{2}>0$ and inverted hierarchy (IH) when $\Delta m_{21}^{2}>0$ and $\Delta m_{31}^{2}<0$.

\begin{table}
\centering
\caption{Possible neutrino mass sum-rule \cite{King}.}
\begin{tabular}{ |p{1cm}||p{1cm}|p{5cm}|p{4cm}|  }
 \hline
 $d$& Type &Neutrino mass sum-rule&Group\\
 \hline
  &1 & $m_{1}+m_{2}=m_{3}$ & $A_{4}, A_{5}, S_{4}, \Delta (54)$\\
 &2 & $m_{1}+m_{3}=2m_{2}$&$S_{4}$\\
1 &3 & $2m_{2}+m_{3}=m_{1}$&$A_{4}, S_{4}, T', T_{7}$\\
 &4 & $m_{1}+m_{2}=2m_{3}$&$S_{4}$\\
 &5 & $m_{1}+\frac{\sqrt{3}+1}{2}m_{3}=\frac{\sqrt{3}-1}{2}m_{2}$&$A'_{5}$\\
\hline
 &6 & $m_{1}^{-1}+m_{2}^{-1}=m_{3}^{-1}$&$A_{4}, S_{4}, A_{5}$\\
 -1 &7 & $2m_{2}^{-1}+m_{3}^{-1}=m_{1}^{-1}$&$A_{4}, T'$\\
&8 & $m_{1}^{-1}+m_{3}^{-1}=2m_{2}^{-1}$&$A_{4}, T'$\\
&9 & $m_{3}^{-1}\pm i m_{2}^{-1}=m_{1}^{-1}$&$\Delta (96)$\\
 \hline
1/2 &10 &$\sqrt{m_{1}}-\sqrt{m_{3}}=2\sqrt{m_{2}}$&$A_{4}\times Z_{2}$\\
&11 &$\sqrt{m_{1}}+\sqrt{m_{3}}=2\sqrt{m_{2}}$&$A_{4}$\\
 \hline
-1/2&12 &$m_{1}^{-1/2}+m_{2}^{-1/2}=2m_{3}^{-1/2}$&$S_{4}$\\
\hline
\end{tabular}
\end{table}

From Tabel 1, we can see that the $A_{4}$ symmetry is the most widely used to describe neutrino mass sum-rule.  Since the $A_{4}$ symmetry is the most widwly used as the underlying symmetry of the neutrino mass sum-rule, therefore we only evaluate the neutrino mass sum-rules that can be described by $A_{4}$ symmetry and other symmetries that can proceed the same mass sum-rule.  Thus, for the next section we only evaluate and discuss neutrino mass sum-rules of type 1 and 3 in Table 1 because those types of neutrino mass sum-rule are the most general mass sum-rule according to the number of underlying symmetries that can be used to describe it.

Meanwhile, neutrinoless double beta decay experiment only give us an upper bound of the effective Majorana mass $\left<m_{ee}\right>$.  The effective Majorana mass is given by
\begin{eqnarray}
\left<m_{ee}\right>=\left|\Sigma V_{ei}^{2}m_{i}\right|,
\end{eqnarray}
where $V_{ei}$ is the {\it i}-th element of the first row of neutrino mixing matrix and $m_{i}$ is the {\it i}-th of the neutrino mass.  

The upper bound of effective Majorana mass which is calculated from neutrinoles double beta decay experiment have been reported by many collaborations.  The Heidelberg-Moscow collaboration \cite{HeMo} which operated five enriched $^{76}$Ge detectors in low-level environment in the Gran Sasso underground laboratoty reported that $\left<m_{ee}\right><0.35$ eV.  The COURICINO collaboration \cite{CUORICINO} reported that they have used $^{130}$Te to detect the neutrinoless double beta decay and the upper bound of the effective Majorana mass is  $\left<m_{ee}\right><0.7$ eV.  In the NEMO3 experiment \cite{NEMO3} the cylindrical source was devided in sectors with enriched $^{100}$Mo and they found that the bound of half life of neutrinoless double beta decay $T_{0\nu}^{1/2}~{\rm (^{100}Mo)}>1.1\times 10^{24}~{\rm y}$ that give the corresponding value of Majorana mass is $\left<m_{ee}\right><1$ eV.

\section{Phenomenological implications of neutrino mass sum-rule}
  As stated in the previous section, we only evaluate the most general neutrino masses in Table 1 (type 1 and 3) prediction on neutrino masses when we use the data of neutrino oscillations as input.  After we know the value of neutrinos masses and its hierarchy, we therefore plot the effective Majorana mass $\left<m_{ee}\right>$ as function of the lightest neutrino mass both for normal and inverted hierarchies.

\subsection{Neutrino mass sum-rule of type 1}
The neutrino mass sum-rule of type 1 reads
\begin{eqnarray}
m_{1}+m_{2}=m_{3}.\label{1}
\end{eqnarray}
From Eq. (\ref{1}) we can have the following relation
\begin{eqnarray}
m_{2}^{2}+2m_{1}m_{2}-\Delta m_{31}^{2}=0,\label{2}
\end{eqnarray}
where $\Delta m_{31}^{2}=m_{3}^{2}-m_{1}^{2}$.  After doing a little algebra, the Eq. (\ref{2}) proceed
\begin{eqnarray}
m_{2}=-2m_{1}+\sqrt{4m_{1}^{2}+\Delta m_{31}^{2}}.\label{3}
\end{eqnarray}

We have also another squared-mass difference 
\begin{eqnarray}
\Delta m_{21}^{2}=m_{2}^{2}-m_{1}^{2},\label{4}
\end{eqnarray}
which can be measured in neutrino oscillation experiment.  By inserting Eq. (\ref{3}) into Eq. (\ref{4}) and solving it to find $m_{1}$, then we have
\begin{eqnarray}
m_{1}=\frac{\sqrt{-105\Delta m_{21}^{2}-15\Delta m_{31}^{2}+60\sqrt{4\Delta m_{21}^{4}-\Delta m_{21}^{2}\Delta m_{31}^{2}+\Delta m_{31}^{4}}}}{15},\label{5}
\end{eqnarray}
or
\begin{eqnarray}
m_{1}=\frac{\sqrt{-105\Delta m_{21}^{2}-15\Delta m_{31}^{2}-60\sqrt{4\Delta m_{21}^{4}-\Delta m_{21}^{2}\Delta m_{31}^{2}+\Delta m_{31}^{4}}}}{15}.\label{6}
\end{eqnarray}

By inserting the central values of squared-mass difference \cite{GG}
\begin{eqnarray}
\Delta m_{21}^{2}=7.59\times 10^{-5}~{\rm eV^{2}},\label{7}\\
\Delta m_{31}^{2}=2.46\times 10^{-3}~{\rm eV^{2}},{\rm for~ NH}\label{8}\\
\Delta m_{31}^{2}=-2.36\times 10^{-3}~{\rm eV^{2}},{\rm for~ IH}\label{9}
\end{eqnarray}
into Eq. (\ref{5}), and Eqs. (\ref{3}) and (\ref{1}), then we have
\begin{eqnarray}
m_{1}=0.021158~{\rm eV},~m_{2}=0.022881~{\rm eV},~m_{3}=0.04404~{\rm eV},
\end{eqnarray}
for normal hierarchy (NH):~$|m_{1}|<|m_{2}|<|m_{3}|$, and
\begin{eqnarray}
m_{1}=0.027615~{\rm eV},~m_{2}=-0.028956~{\rm eV},~m_{3}=-0.001342~{\rm eV},
\end{eqnarray}
for inverted hierarchy (IH):~$|m_{3}|<|m_{1}|<|m_{2}|$.

If we use Eq. (\ref{6}) to determine $m_{1}$, $m_{2}$ from Eq. (\ref{3}), and $m_{3}$ from Eq. (\ref{1}) then we have the hierarchy of neutrino masses as follow
\begin{eqnarray}
|m_{1}|<|m_{3}|<|m_{2}|,\\
\end{eqnarray}
or
\begin{eqnarray}
|m_{2}|<|m_{2}|<|m_{1}|,
\end{eqnarray}
when $\Delta m_{31}^{2}>0$, and
\begin{eqnarray}
|m_{2}|<|m_{1}|<|m_{3}|,\\
\end{eqnarray}
or
\begin{eqnarray}
|m_{1}|<|m_{3}|<|m_{2}|,
\end{eqnarray}
when $\Delta m_{31}^{2}<0$.  Thus, only neutrino mass of Eq. (\ref{5}) with neutrino mass sum-rule of type 1 can predict the hierarchy of neutrino mass in agreement with the experimental data of neutrino oscillations.  Plot of effective Majorana mass as function of the lightest neutrino mass for neutrino mass sum rule of type 1  with the mixing angles $\theta_{12}=37^{\circ}$ and $\theta_{13}=5^{\circ}$ and the central value of the squared-mass differences in Ref. \cite{GG} are used as input(red line for NH and green line for IH) is displayed in Figure 1.

\begin{figure}[h]
\centering
\includegraphics[width=0.50\textwidth]{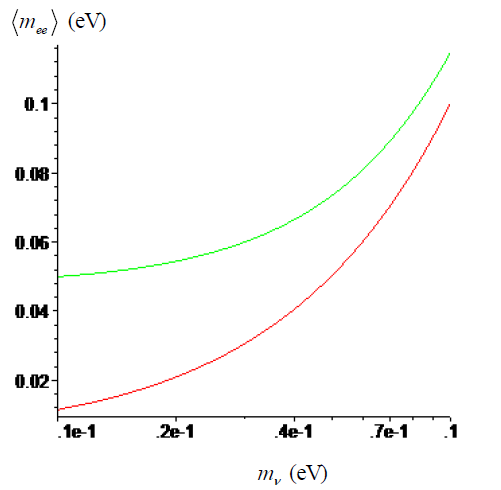}
\caption{Plot of $\left<m_{ee}\right>$ as function of lightest $m_{\nu}$ for sum-rule of type 1 }
\end{figure}

\subsection{Neutrino mass sum-rule oftType 3}
As shown in Table 1, the neutrino mass sum-rule of type 3 reads
\begin{eqnarray}
2m_{2}+m_{3}=m_{1}.\label{10}
\end{eqnarray}
From neutrino mass sum-rule of Eq. (\ref{10}) we can have
\begin{eqnarray}
4m_{2}^{2}+4m_{3}m_{2}+\Delta m_{31}^{2}=0,
\end{eqnarray}
which then proceed
\begin{eqnarray}
m_{2}=-\frac{\Delta m_{31}^{2}-4\Delta m_{32}^{2}}{2\sqrt{-\Delta m_{31}^{2}+4\Delta m_{32}^{2}}},\label{11}
\end{eqnarray}
or
\begin{eqnarray}
m_{2}=\frac{\Delta m_{31}^{2}-4\Delta m_{32}^{2}}{2\sqrt{-\Delta m_{31}^{2}+4\Delta m_{32}^{2}}}.\label{12}
\end{eqnarray}

It is apparent from Eqs. (\ref{11}) and (\ref{12}) that neutrino mass $m_{2}$ is only as function of squared-mass differences $\Delta m_{31}^{2}$ and $\Delta m_{32}^{2}$.  Since we need squared-mass difference $m_{32}^{2}$ to find out the value of $m_{2}$, then we can use the advantage of defenition squared-mass differences $m_{21}^{2}$ and $m_{31}^{2}$ which then proceeds
\begin{eqnarray} 
\Delta m_{32}^{2}=\Delta m_{31}^{2}-\Delta m_{21}^{2}.\label{13}
\end{eqnarray}

By applying the same above procedure in determining the neutrino masses, from Eq. (\ref{11}) we have neutrino masses for NH as follow
\begin{eqnarray}
m_{1}=0.015869~{\rm eV},~m_{2}=0.018103~{\rm eV},~m_{3}=0.052075~{\rm eV},
\end{eqnarray}
and when using Eq. (\ref{12}) we have
\begin{eqnarray}
m_{1}=0.015869~{\rm eV},~m_{2}=0.033972~{\rm eV},~m_{3}=-0.052075~{\rm eV},
\end{eqnarray}
which is consistent with normal hierarchy:$|m_{1}|<|m_{2}|<|m_{3}|$. 

Meanwhile, for IH, by using Eq. (\ref{12}) we have neutrino masses
\begin{eqnarray}
m_{1}=0.018790i~{\rm eV},~m_{2}=-0.016648i~{\rm eV},~m_{3}=0.052087i~{\rm eV},\label{14}
\end{eqnarray}
and when using Eq. (\ref{12}) we have
\begin{eqnarray}
m_{1}=0.018790i~{\rm eV},~m_{2}=0.035439i~{\rm eV},~m_{3}=-0.052087i~{\rm eV}.\label{15}
\end{eqnarray}
Both the obtained neutrino masses in Eqs. (\ref{14}) and (\ref{15}) are incosistent with the inverted hierarchy.  Thus, we can olnly use the neutrino mass $m_{2}$ of Eq. (\ref{11}) and neutrino mass sum-rule of type 3 to predict the correct hierarchy of neutrino mass.  Plot of effective Majorana mass as function of the lightest neutrino mass for neutrino mass sum rule of type 3  with the mixing angles $\theta_{12}=37^{\circ}$ and $\theta_{13}=5^{\circ}$ and the central value of the squared-mass differences in Ref. \cite{GG} are used as input (only allowed NH) is displayed in Figure 2.

\begin{figure}[h]
\centering
\includegraphics[width=0.60\textwidth]{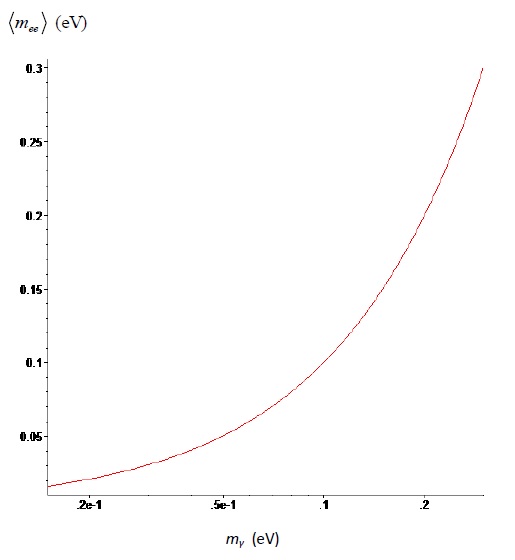}
\caption{Plot of $\left<m_{ee}\right>$ as function of $m_{\nu}$ for sum-rule of type 3}
\end{figure}

We can see from the two types of neutrino mass sum-rule that the effective Majorana mass depend on the character of neutrino mass spectrum.  In \cite{Bilenky} another mass hierarchy i. e. quasi-degenerate hierarchy, the expected value of the effective Majorana mass is relatively large which partly excluded by the data of the performed double beta decay experiments and cosmological data.

\section{Conclusions}
We have briefly review neutrino mass sum-rule that can be read in lietratures and we found there 12 type of neutrino mass sum-rule that can be derived from various symmetries.  The most widely symmetry is $A_{4}$ symmetry.  Based on the widely used symmetry that can be applied to derive the neutrino mass sum-rule, we choose two tyoes neutrino mass sum-rule that have already been reported in literature i. e. type 1 and type 3.  When we evaluate the predictions of those both neutrino mass sum-rules on neutrino masses and its hierarchy by using the advantages of neutrino oiscillations data as input, we find that the neutrino mass sum-rule of type 1 can predict neutrino mass hierarchy both in normal hierarchy and inverted hierarcy.  Meanwhile, the neutrino mass sum-rule of type 3 can only predict neutrino mass hierarchy in normal hierarchy.

\section*{Acknowledgement}
This researh is partly supported by Hibah Penelitian Internal LPPM USD with contract No. 070/Penel./LPPM-USD/IV/2017.

\end{document}